\documentclass[preprint,amsmath,amssymb,prb,longbibliography]{revtex4-1}

\usepackage{amssymb}
\usepackage{amsfonts}
\usepackage{color}
\usepackage{dsfont}
\usepackage[normalem]{ulem}
\usepackage{dcolumn}
\usepackage{mathtools}
\usepackage{siunitx}
\usepackage[colorlinks=true,linkcolor=blue,citecolor=blue,urlcolor=blue]{hyperref}

\usepackage{graphicx}
\usepackage{blindtext}
\usepackage{outlines}

\usepackage{enumitem}
\setlist[enumerate,2]{label=\roman*)}
\setlist[enumerate,3]{label=\alph*)}

\newcommand{\unitmatr}{\ensuremath{\mathds{1}}}

\newcommand{\dd}{\ensuremath{\text{d}}}

\newcommand{\iu}{\mathrm{i}}

\newcolumntype{d}[1]{D{.}{.}{#1}}

\definecolor{orange}{rgb}{1,0.5,0}
\definecolor{darkgreen}{RGB}{0,100,0}

\makeatother

\begin{document}

\title{Impurity-induced orbital magnetization in a Rashba electron gas}
\author{Juba Bouaziz}
\email{j.bouaziz@fz-juelich.de}
\author{Manuel dos Santos Dias}
\author{Filipe Souza Mendes Guimar\~aes}
\author{Stefan Bl\"ugel}
\author{Samir Lounis}
\affiliation{Peter Gr\"unberg Institut and Institute for Advanced Simulation, 
Forschungszentrum  J\"ulich and JARA, 52425 J\"ulich, Germany}

\date{\today}

\begin{abstract}
We investigate the induced orbital magnetization density in a Rashba electron 
gas with magnetic impurities. Relying on classical electrodynamics, we obtain 
this quantity through the bound currents composed of a paramagnetic and a 
diamagnetic-like contribution which emerge from the spin-orbit interaction.
Similar to Friedel charge ripples, the bound currents and the orbital magnetization density oscillate as function of distance away from the impurity with characteristic wavelengths defined by the Fermi energy and the strength of the Rashba spin-orbit interaction. The net induced orbital magnetization was found to be of the order of magnitude of its spin counterpart. 
Besides the exploration of the impact of the electronic filling of the impurity states, we investigate and analyze the orbital magnetization induced by an equilateral frustrated trimer in various non-collinear magnetic states. On the one hand, we confirm that non-vanishing three-spin chiralities generate a finite orbital magnetization density.
On the other hand, higher order contributions lead to multiple-spin chiralities affecting non-trivially and significantly the overall magnitude and sign of the orbital magnetization.
\end{abstract}

\maketitle

\section{Introduction}
The inversion symmetry breaking at surfaces and interfaces leads to the 
emergence of a wide variety of phenomena. Its signature can be 
detected experimentally through different physical quantities~\cite{Hellman:2017,Sinova:2015,Manchon:2015}. 
In combination with the spin-orbit (SO) interaction, it leads 
to the Rashba effect, which consists of an energy spin-splitting of 
the surface/interface states~\cite{Rashba:1960,Bychkov:1984,Manchon:2015}. 
The Rashba effect was observed experimentally at noble metallic surfaces 
using angle-resolved photoemission spectroscopy (ARPES)~\cite{Lashell:1996,Reinert:2001,Nicolay:2001} 
through the energy dispersion imaging of the Rashba spin-splitting. 
Since its first observations, several studies were devoted to this 
effect as reported in several reviews~\cite{Manchon:2015,Bihlmayer:2015}. For instance, it was shown 
that the Rashba spin-splitting can be manipulated by material engineering~\cite{Gierz:2009,Liebmann:2015,Elmers:2016}. 
In the case of a Bi monolayer deposited on Si(111), the splitting can 
be very large~\cite{Gierz:2009}.

An alternative way to probe the spin-splitting of the surface states is using the 
Friedel oscillations~\cite{Friedel:1958} resulting from the scattering of the surface 
electrons off defects or impurities. They can be accessed experimentally via scanning 
tunneling microscopy (STM)~\cite{Crommie:1993,Fiete:2003}. However, it was shown
theoretically that no signature of the Rashba spin-splitting can be observed in the
charge density surrounding a single impurity~\cite{Petersen:2000}. In Ref~\onlinecite{Lounis:2012}, 
Lounis {\it et al.} showed that the introduction of a magnetic impurity into a 
Rashba electron gas causes Friedel oscillations in the spin magnetization density 
where a signature of the spin-splitting can be observed --- the induced spin magnetization 
exhibits a Skyrmion-like spin texture. 

In addition to the induced spin magnetization, the presence of the Rashba SO 
interaction in conjunction with an external magnetic field/moment (breaking time-reversal 
symmetry) generates bound charge currents. They were explored in the context of 
magnetic impurities and ferromagnetic islands coupled to superconductors with finite 
spin-orbit interaction~\cite{Pershoguba:2015}. These bound currents represent electrons 
moving in a closed circuit and produce a finite orbital magnetization within the Rashba 
electron gas, which can be of the order of magnitude of the spin magnetization.
The time-reversal invariant Rashba electron gas can also be seen as having a 
compensated orbital magnetization~\cite{Dyrda:2016}.

In classical electrodynamics, the bound currents and orbital magnetization are 
related via~\cite{THONHAUSER:2011}:
\begin{equation}
\vec{j}(\vec{r}) = \vec{\nabla}_{\vec{r}}\times\vec{m}_{l}(\vec{r})\quad, 
\label{bc_orb_magn}
\end{equation}
where $\vec{m}_{l}(\vec{r})$ is the orbital magnetization density and $\vec{j}(\vec{r})$ 
is the bound current density. In equilibrium, $\vec{j}(\vec{r})$ is non-dissipative and 
fulfills the continuity equation for the electron density $\rho(\vec{r})$ (see 
Eq.~\eqref{continuity_eq}). Furthermore, the magnetic impurities also generate finite 
ground state spin currents leading to the emergence of the chiral Dzyaloshinskii-Moriya (DM) interaction~\cite{Kikuchi:2016,Koretsune:2018,Freimuth:2017}. The latter is the key 
ingredient for the stabilization of topological spin textures such as magnetic 
Skyrmions~\cite{Bogdanov:1989,Nagaosa:2013,Fert:2017,Fert:2013}. Recently, it was shown that for 
such entities, a finite orbital magnetization emerges even in absence of the spin-orbit 
interaction. In this case, it arises from the scalar spin chirality of the spin texture, 
$\mathcal{C}_{ijk} = \vec{S}_{i}\cdot\big(\vec{S}_{j}\times\vec{S}_{k}\big)$, 
with $\vec{S}_{i}$ being the spin magnetic moment direction at site $i$~\cite{Manuel:2016,Manuel:2017}. 
Indeed, non-collinear spin textures can be viewed as a gauge field that couples 
the spin and orbital degrees of freedom~\cite{Xiao:2010,Karin:2014}, mimicking the 
effects of the SO interaction. Moreover, for large magnetic Skyrmions, the chirality-driven 
orbital magnetization is quantized and becomes topological (\textit{i.e.} not affected 
by continuous deformations of the magnetic texture) and might be accessed experimentally 
via x-ray magnetic circular dichroism (XMCD)~\cite{Manuel:2016}. The chiral orbital 
magnetization was also found in periodic systems~\cite{Hoffman:2015,Hanke:2016,Hanke:2017}
as well as in continuous topological structures deposited on Rashba electron gas~\cite{Fabian:2018}. 
Finally, it was shown in Ref.~\onlinecite{Tatara:2003} that non-vanishing induced 
bound currents arise within a two dimensional electron gas when magnetic moments 
possessing a non-zero spin chirality are deposited on top. 

In this manuscript, we investigate the induced orbital magnetization generated when 
magnetic impurities are deposited on a Rashba electron gas considering a single impurity 
or a trimer with a frustrated spin state, and show that it can be of the order of magnitude of the spin 
magnetization. Furthermore, we demonstrate that higher order spin chiralities can provide 
a substantial contribution to the induced orbital magnetization generated by 
clusters involving more than one impurity. The paper is structured 
as follows: First, we discuss the bound currents and their different contributions 
(paramagnetic and diamagnetic), which are evaluated analytically for the single 
impurity case. Second, the induced orbital magnetization density is computed 
starting from the bound currents by numerically solving a Poisson equation. 
The impact of the impurity nature on the orbital magnetization 
by tuning the scattering phase shifts is also considered. Finally, we compute the 
orbital magnetization for a magnetic trimer in an equilateral triangle with and 
without the SO interaction, and provide functional forms connecting the spin impurity 
moments to the orbital magnetization. 

\section{Rashba Model}

The SO interaction leads, in a structure-asymmetric environment such as surfaces 
or interfaces, to a spin-splitting of the degenerate eigenstates for the two-dimensional 
free electron gas. The model of Bychkov and Rashba~\cite{Rashba:1960,Bychkov:1984} 
describes this splitting by adding a linear term in momentum $\vec{p}$ to the kinetic 
energy of the free electrons. The so-called Rashba Hamiltonian is given by
\begin{equation}
\boldsymbol{H}_\text{R} = \frac{ p^2_{x} + p^2_{y} }{2m^*}\,\unitmatr_2- 
\frac{\alpha_\text{so}}{\hbar}(\boldsymbol{\sigma}_{x} {p_{y}} - 
\boldsymbol{\sigma}_{y} {p_{x}})\quad,
\label{rashba_hamiltonian}
\end{equation}
where $\{p_x,p_y\}$ are the components of the momentum operator $\vec{p}$ 
in Cartesian coordinates of the surface plane whose normal points along 
$\vec{e}_z$, and $m^*$ is the effective mass of the electron. $\boldsymbol{\sigma}_x$ 
and $\boldsymbol{\sigma}_y$ are Pauli matrices and $\unitmatr_2$ is the 
unit matrix in spin-space with a global spin frame of reference parallel 
to the $z$-axis. $\alpha_\text{so}$ is known as Rashba parameter and represents 
the strength of the SO interaction. The linear term in Eq.~\eqref{rashba_hamiltonian} 
is induced by a SO gauge field given by~\cite{Hatano:2007,SonHsien:2008}
\begin{equation}
\vec{\boldsymbol{A}}_\text{so} = \frac{m^{*}\, \alpha_\text{so}}{e\hbar}\ 
(-\boldsymbol{\sigma}_{y}, \boldsymbol{\sigma}_{x})\quad,
\end{equation}
where $e$ is the electron charge. Using this SO gauge field, 
the Hamiltonian is expressed as
\begin{equation}
\boldsymbol{H}_\text{R} = \frac{\big(\vec{p}-e\vec{\boldsymbol{A}}_\text{so}\big)^2}{2m^{*}} 
- V_\text{so}\quad,
\end{equation}
with $V_\text{so} = \frac{m^*\alpha_\text{so}^{2}}{\hbar^2}$ being a constant. 
Since $[\boldsymbol{A}^{x}_\text{so},\boldsymbol{A}^{y}_\text{so}] \ne 0$, 
$\vec{\boldsymbol{A}}_\text{so}$ is a non-Abelian gauge field, which complicates 
any possible approach relying on gauge transformations~\cite{SonHsien:2008}.
Starting from the time-dependent Schr\"odinger equation, $\iu\hbar \frac{\partial 
\psi(\vec{r},t)}{\partial t} = \boldsymbol{H}_\text{R} \,\psi(\vec{r},t)$, 
we arrive at a continuity equation for the electron charge density $\rho(\vec{r},t) 
= |\psi(\vec{r},t)|^2$ that provides the current density $\vec{j}(\vec{r},t)$ 
and the current operator ${\vec{\boldsymbol{j}}}$,
\begin{equation}
\frac{\partial \rho(\vec{r},t)}{\partial t} + 
\vec{\nabla}_{\vec{r}}\cdot\vec{j}(\vec{r},t) = 0\quad,
\label{continuity_eq}
\end{equation}
\begin{equation}
{\vec{\boldsymbol{j}}} = \frac{\hbar}{2m^{*}\iu}\, \lim_{\vec{r}^{\,\prime} 
\rightarrow \vec{r}}(\vec{\nabla}_{\vec{r}} - 
\vec{\nabla}_{{\vec{r}}^{\,\prime}})\,\unitmatr_2 - \frac{e}{m^*}\, 
 \vec{\boldsymbol{A}}_\text{so}\quad.
 \label{current_op}
\end{equation}
The first term in Eq.~\eqref{current_op} is the paramagnetic contribution to the 
current operator, while the second term is a diamagnetic-like contribution arising 
from the SO gauge field. Both parts are included in the calculations presented in this manuscript.
Furthermore, when starting from the Dirac Hamiltonian and performing an 
expansion in the non-relativistic limit, one finds an extra 
contribution to the current operator coming from the Zeeman term of the Hamiltonian, 
${\vec{\boldsymbol{j}}}_\text{Zeeman} = \frac{\hbar}{2m^{*}}\,\lim_{\vec{r}^{\,\prime} 
\rightarrow \vec{r}}\vec{\nabla}_{\vec{r}}\times\vec{\boldsymbol{\sigma}}$. It may 
be induced either by a magnetic field or a finite magnetization~\cite{Berche:2013}. 
However, this term is not included in our discussion since it does not contribute 
to the orbital magnetization which is the quantity of interest in this manuscript.

\section{Bound currents emerging from a single magnetic impurity on a Rashba electron gas}
\label{Bound_current_tmat}

The introduction of magnetic impurities into the system leads to the breaking of 
time-reversal symmetry, which in presence of SO interaction is expected to induce 
a finite orbital magnetization~\cite{THONHAUSER:2011}. The magnetic impurities 
are embedded into the Rashba electron gas using a Green function approach in real 
space via the Dyson equation
\begin{equation}
\begin{split}
\boldsymbol{G}(\vec{r},\vec{r}^{\,\prime},\varepsilon) & = 
\, \boldsymbol{G}^\text{R}(\vec{r},\vec{r}^{\,\prime},\varepsilon)\\
& + \sum_{i j} \boldsymbol{G}^\text{R}(\vec{r},\vec{r_{i}},\varepsilon)
\,\boldsymbol{\tau}_{ij}(\varepsilon)\, \boldsymbol{G}^\text{R}
(\vec{r_{j}},\vec{r}^{\,\prime},\varepsilon)\quad
\end{split}
\label{dyson_eq_gf}
\end{equation}
relating the Green function of the Rashba electron gas $\boldsymbol{G}^\text{R}
(\vec{r},\vec{r}^{\,\prime},\varepsilon)$ (see Appendix A for an explicit expression)
to the Green function of the Rashba electron gas with impurities $\boldsymbol{G}
(\vec{r},\vec{r}^{\,\prime},\varepsilon)$ through the scattering path operators 
$\boldsymbol{\tau}_{ij}(\varepsilon)$ ($i,j$ running over the impurities). The 
latter describes single and multiple scattering processes experienced by the 
electrons at the impurities and can be computed from the transition matrices for isolated impurities
(t-matrix) $\boldsymbol{t}_{i}(\varepsilon)$ as
\begin{equation}
\begin{split}
\boldsymbol{\tau}_{ij}(\varepsilon) & = \boldsymbol{t}_{i}(\varepsilon)
\,\delta_{ij} \\ & + \sum_{k} \boldsymbol{t}_{i}(\varepsilon)
\,\boldsymbol{G}_{ik}^\text{R}(\varepsilon)\,(1-\delta_{ik})\ 
\boldsymbol{\tau}_{kj}(\varepsilon)\quad.
\end{split}
\label{scatter_path}
\end{equation}
Furthermore, considering that the Fermi wave length of the Rashba electrons is 
much larger compared to the spatial extension of the impurities, we employ the 
s-wave approximation~\cite{Fiete:2003}. In this case, the transition matrix is 
diagonal in spin space, $t^{\sigma}_{i}(\varepsilon) = \frac{\iu\hbar}{m^*}
(e^{2\iu\delta_{i}^{\sigma\sigma}(\varepsilon)}-1)$, with $\delta_{i}^{\sigma}(\varepsilon)$ 
representing the scattering phase shift, by choosing the magnetic moment of 
the impurity along the z-axis.

For an Fe impurity, it can be approximated by $\delta_{i}^{\uparrow}= \pi$ and $\delta_{i}^{\downarrow}=\frac{\pi}{2}$~\cite{Juba:2017}. For the single impurity 
case, Eq.~\eqref{scatter_path} reduces to $\boldsymbol{\tau}_{ij}(\varepsilon) 
= \boldsymbol{t}_{i}(\varepsilon)\,\delta_{ij}$ and $\vec{j}(\vec{r})$ can be 
computed analytically. For an impurity located at the origin with its magnetic 
moment pointing perpendicular to the surface (\textit{i.e.} along the $z$-axis), 
the cylindrical symmetry of the Rashba electron gas is preserved and the current 
density in cylindrical coordinates $\vec{r} = (r\cos\theta,r\sin\theta)$ reads 
\begin{equation}
\begin{split}
\vec{j}(\vec{r}) = -\frac{\hbar}{m^*\pi}\,\text{Im}\int_{0}^{\varepsilon_\text{F}}
\dd\varepsilon\, &\Big{[}\frac{G^{2}_\text{ND}(r,\varepsilon)}{r} 
-\frac{2m^*\alpha_\text{so}}{\hbar^{2}} 
G_\text{D}(r,\varepsilon) G_\text{ND}(r,\varepsilon) \Big{]}\\
&\,\Delta{t}_{i}(\varepsilon)\,{\vec{e}_{\theta}}\quad.
\end{split}
\label{current_mz}
\end{equation}
${\vec{e}_{\theta}} = (\sin\theta,-\cos\theta)$ is the azimuthal unit vector 
at point $\vec{r}$. $G_\text{D}(r,\varepsilon)$ and $G_\text{ND}(r,\varepsilon)$ 
are functions of the distance $r=|\vec{r}|$ and energy $\varepsilon$, and 
represent the diagonal and off-diagonal parts of the Rashba Green function 
in spin space, respectively. More details on this derivation are given 
in Appendix~\ref{sec:Appendix_A}. 

A comment concerning the energy integration in Eq.~\eqref{current_mz} is in order. 
The integration is only performed from $[0,\varepsilon_\text{F}]$. 
The energy range $[-\varepsilon_\text{R},0]$ with $\varepsilon_\text{R} = 
\frac{m^*\alpha^{2}_\text{so}}{2\hbar^{2}}$ being the Rashba energy 
is not included, for two reasons. First, for realistic values of the 
Rashba parameter we have $\varepsilon_\text{R}\ll\varepsilon_\text{F}$, 
and so this energy range is very small. Second, although this energy 
range contains a Van Hove singularity, a careful analysis shows that
that the t-matrix cancels the singularity and leads to a smooth energy 
dependence of the Green function~\cite{Pershoguba:2016,Juba:2017}. 
Combining both arguments, we conclude that one can safely neglect the 
contribution from this energy range. 

The first term in Eq.~\eqref{current_mz} represents the paramagnetic part of the 
current density, while the second is the diamagnetic one. $\vec{j}(\vec{r})$ has 
no radial component, thus swirling around the magnetic impurity. 
The dependence of the current on the intrinsic properties of the impurity is entirely 
encoded in $\Delta{t}_{i}(\varepsilon) = t^{\uparrow\uparrow}_{i}(\varepsilon) - 
t^{\downarrow\downarrow}_{i}(\varepsilon)$. This result reveals, in a clear fashion, 
that a finite orbital magnetization requires a spin magnetization/magnetic field 
breaking time-reversal symmetry ({\it i.e.} $t^{\uparrow \uparrow}_{i}(\varepsilon) 
\ne t^{\downarrow \downarrow}_{i}(\varepsilon)$) and a broken space inversion symmetry 
environment with the SO interaction $({\it i.e}\ G_\text{ND}(r,\varepsilon) \ne 0)$. 
Similar results were obtained for magnetic impurities deposited on superconductors 
with Rashba spin-orbit interaction~\cite{Pershoguba:2015}.

\begin{figure*}
\hspace{-4mm}
\includegraphics[width=1.0\textwidth]{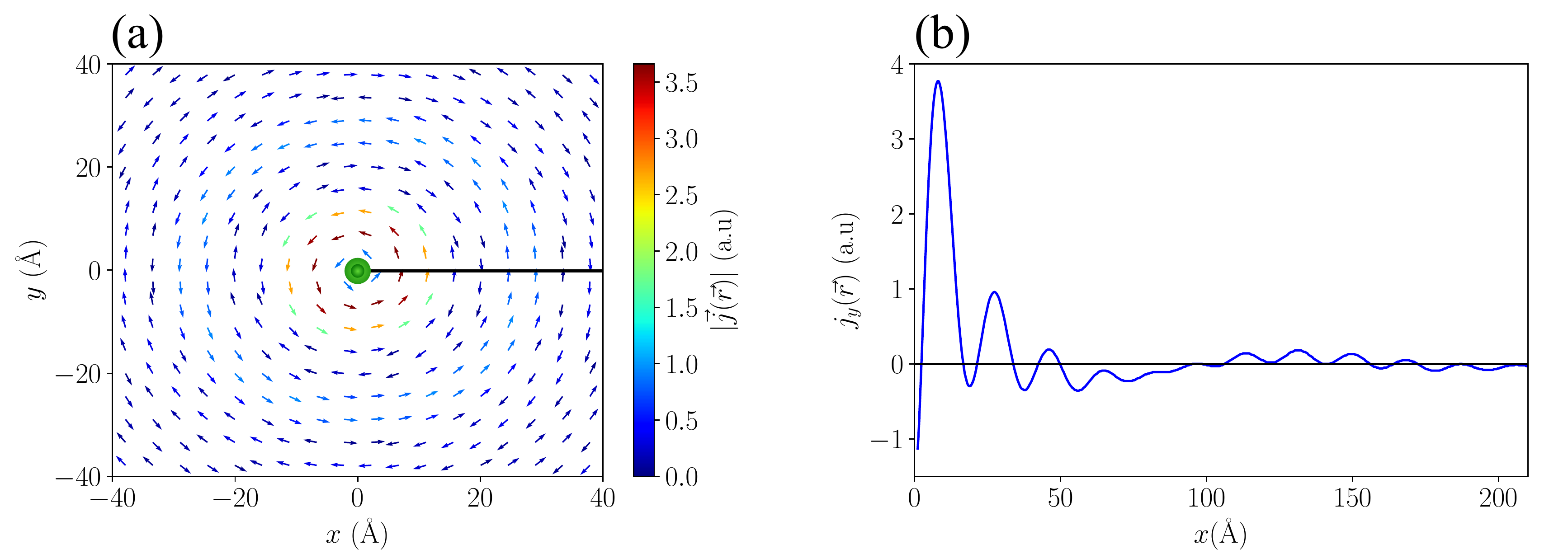}
 \caption{Ground state charge currents induced by a single Fe impurity on a 
 	Au(111) with a magnetic moment perpendicular to surface plane (along the 
 	$z$-axis). $\alpha_\text{so}=\SI{-0.4}{\electronvolt\angstrom}$, 
 	$m^*=0.26\,m_\mathrm{e}$ and $\varepsilon_\text{F} = 410$ meV~\cite{Walls:2006} are the 
 	Rashba model parameters for the Au(111) surface state. 
    The Fe impurity is considered in the s-wave approximation~\cite{Juba:2017}. a) The 
    dissipationless currents are swirling around the magnetic impurity in agreement 
    with the continuity equation and the axial symmetry of the system. b) Evolution of the $y$-component of the current 
    density as function of the distance from the impurity. It displays an 
    oscillatory behaviour with two wave lengths $\lambda_\text{F}\sim\SI{18.5}{\angstrom}$
    and $\lambda_\text{so}\sim\SI{130}{\angstrom}$.} 
\label{charge_curr_map}
\end{figure*}
In Fig.~\ref{charge_curr_map}a, we show the ground state charge currents induced 
by a single Fe impurity deposited on the Rashba surface states of a Au(111) surface, 
computed from Eq.~\eqref{current_mz}. The Rashba model parameters are 
$\alpha_\text{so}=\SI{-0.4}{\electronvolt\angstrom}$, $m^*=0.26\,m_\mathrm{e}$ 
($m_\text{e}$ being the electron mass) 
and $\varepsilon_\text{F} = 410$ meV~\cite{Walls:2006}. These swirling bound 
currents are dissipationless (\textit{i.e.} with zero divergence) with an 
oscillating amplitude reminiscent of the Friedel oscillations present in the 
charge and spin densities. A cut at $y=0$ is shown in Fig.~\ref{charge_curr_map}b, 
where the oscillating current density displays a beating effect at $x\sim60$\,\AA\,similar to 
the one observed in the spin magnetization density and magnetic exchange interactions 
characterizing single impurities embedded in a Rashba electron gas~\cite{Lounis:2012,Juba:2017}.
Two wavelengths are at play in settling the oscillatory behavior of the current density: a short one given 
by the Fermi wave length $\lambda_\text{F}\sim \SI{18.5}{\angstrom}$ and a 
long one induced by the SO interaction $\lambda_\text{so}\sim\SI{130}{\angstrom}$. 
This behavior can be understood when considering the analytical form of the 
current density obtained in the asymptotic limit (\textit{i.e.} expanding 
$G_\text{D}(r,\varepsilon)$ and $G_\text{ND}(r,\varepsilon)$ for $r\rightarrow\infty$):
\begin{equation}
\begin{split}
\vec{j}({r}) & = -\frac{m^{*}}{\hbar^{3}\pi r}\,\text{Im}\,\Delta{t}_{i}\left[6k^{2}_\text{so}\left[\text{CI}(2k_\text{F}r)-\text{CI}(2|k_\text{so}|r)\right]
- 2k^{2}_\text{so}\left[\frac{\sin (2k_\text{F}r)}{2k_\text{F}r}-\frac{\sin (2|k_\text{so}|r)}{2|k_\text{so}|r}\right]\right]\vec{e}_{\theta}\\
&  +\frac{2(m^{*})^{2}\alpha_\text{so}}{\hbar^{4}\pi}\,\text{Im}\,\Delta{t}_{i}\left[ 4k^{2}_\text{so}\left[\text{SI}(2k_\text{F}r)-\text{SI}(2|k_\text{so}|r)\right] + \frac{1}{r^{2}}\left[\sin(2k_\text{F}r)-\sin(2|k_\text{so}|r)\right]\right]\vec{e}_{\theta}\quad.
\end{split}
\label{analytical_current_den}
\end{equation}
$\text{CI}(x)$ and $\text{SI}(x)$ represent the sine and cosine integrated functions of $x$,
while $k_\text{so} = \frac{m^{*}\alpha_\text{so}}{\hbar^{2}}$. Eq.~\eqref{analytical_current_den} 
shows that in the asymptotic limit, the current density oscillates 
with two different wave lengths $\lambda_\text{F} = \frac{\pi}{k_\text{F}}$
and $\lambda_\text{so}\propto\frac{\pi}{k_\text{so}}$.

\section{Method for the evaluation of the orbital magnetization}
\label{gs_charge_current_morb}

In the previous section, we showed that a system with broken time and space inversion 
symmetry hosts ground state charge currents. These currents give rise to a finite orbital 
magnetization within the Rashba electron gas. In absence of free charge currents and 
time-dependent external fields, the orbital magnetization density $\vec{m}_{l}(\vec{r})$ 
is related to the ground state charge current via Eq.~\eqref{bc_orb_magn}. 
Let us begin by showing that there is no indeterminacy in this relation, contrary to 
the three-dimensional case~\cite{Hirst:1997}. Due to the two-dimensional geometry,
the current density lies in $xy$-plane, and so the orbital magnetization is restricted 
to the $z$-direction. The standard indeterminacy in Eq.~\eqref{bc_orb_magn} consists of 
adding the gradient of an arbitrary function to the orbital magnetization, 
which does not affect the current density. Since only the $z$-component of the 
gradient is compatible with this geometry and it vanishes identically, there is 
no remaining freedom in the definition of the orbital magnetization density. 
Therefore, we can use Eq.~\eqref{bc_orb_magn} to define the orbital magnetization
density, by rewriting it as a Poisson equation,
\begin{equation}
\begin{split}
\vec{\nabla}_{\vec{r}}\times\vec{j}(\vec{r}) & = \vec{\nabla}_{\vec{r}}
\times\vec{\nabla}_{\vec{r}}\times\vec{m}_{l}(\vec{r})\quad,\\
& = \vec{\nabla}_{\vec{r}}\,(\vec{\nabla}_{\vec{r}}\cdot\vec{m}_{l}(\vec{r})) 
- \vec{\nabla}^{2}_{\vec{r}}\cdot\vec{m}_{l}(\vec{r})\quad.
\end{split}
\label{poisson_magn}
\end{equation}
For 2D systems, the previous equation reduces to: 
\begin{equation}
\partial_{x}\,j_{y}(\vec{r}) -\partial_{y}\,j_{x}(\vec{r})  = - 
\vec{\nabla}^{2}_{\vec{r}}\,{m}_{l,z}(\vec{r})\quad,
\label{poisson_magn2}
\end{equation}
which can be solved numerically using a Fourier series in a large finite simulation box.
The Fourier components $m_{l,z}(\vec{k})$ of the orbital magnetization  are
\begin{equation}
m_{l,z}(\vec{k}) = \iu\,\frac{k_{y}\,j_{x}(\vec{k})-k_{x}\,j_{y}(\vec{k})}{k^{2}_{x}+k^{2}_{y}}\quad,
\label{fourier_transform_morb}
\end{equation}
where $j_{\alpha}(\vec{k})$ is the Fourier transform of $j_{\alpha}(\vec{r})$ defined as:
\begin{equation}
j_{\alpha}(\vec{k}) = \sum_{i=1}^{N_\text{r}} j_{\alpha}(\vec{r}_{i})\,e^{\iu\vec{k}\cdot\vec{r}_{i}}\quad.
\label{ft_current}
\end{equation}
In practice, we consider Fe impurities deposited on Au(111)~\cite{Juba:2017}, in a box of $\SI{210}{\angstrom}\times\SI{210}{\angstrom}$ divided in a grid of $N_\text{r} = 1000\times1000$ 
points in real space. In our calculations, the Fermi wave length is set to $\lambda_\text{F}\sim\SI{18.5}
{\angstrom}$. The ratio between $\lambda_\text{F}$ and the grid spacing is thus $0.01$, which 
was found to lead to converged results. This large box also ensures that $\vec{j}(\vec{r}) \simeq 0$ 
at the edges of the box to avoid interactions between the impurity and its periodic copies. 
Lastly, $m_{l,z}(\vec{k})$ is Fourier transformed back to real space providing ${m}_{l,z}(\vec{r})$. 

\section{Orbital magnetization induced by a single impurity}

\begin{figure*}
	\centering
	\includegraphics[width=1.0\textwidth]{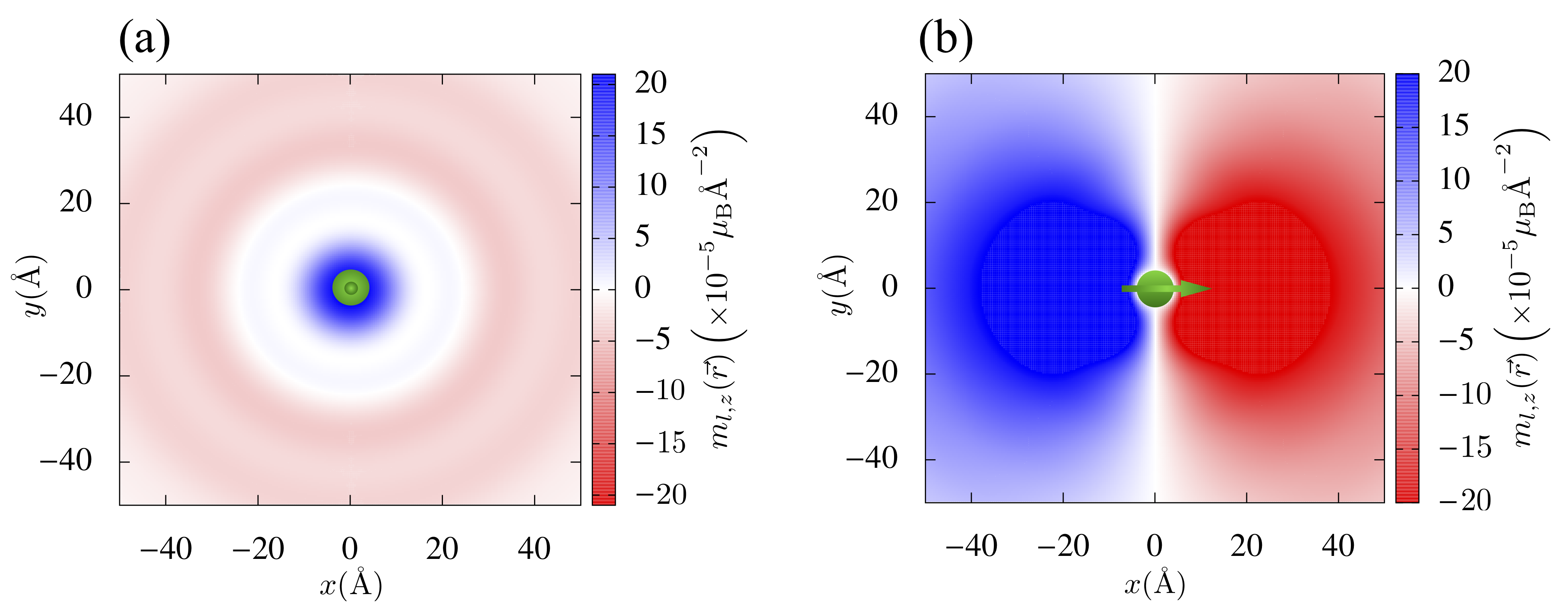}
	\caption{Induced orbital magnetization map for a single Fe adatom deposited on 
		a Rashba electron gas using the same model parameters as in Sec.~\ref{Bound_current_tmat}. 
		The Fe impurity taken in the s-wave approximation~\cite{Fiete:2003} is 
		represented by a green sphere located at the origin, while its magnetic 
		moment is represented by a green arrow. 
		(a) When the impurity spin moment points along the $z$-direction, the 
		induced orbital magnetization consists of concentric rings centered around 
		the Fe impurity oscillating with two characteristic wave lengths 
		$\lambda_\text{F}$ and $\lambda_\text{so}$. (b) When the impurity spin 
		moment lies in the plane, along the $x$-direction, the orbital magnetization 
		is strongly anisotropic since $m^{z}_{l}$ is positive (negative) for 
		negative (positive) $x$.} 
	\label{orb_magn_mz}
\end{figure*}

We now discuss the orbital magnetization obtained for the setup discussed in 
Sec.~\ref{gs_charge_current_morb} in presence of a single magnetic impurity with a 
spin moment pointing perpendicular to the plane ($z$-axis), and also when it points 
in the plane along the $x$-axis. Their orbital magnetization densities obtained using 
Eq.~\eqref{poisson_magn2} are shown in Figs.~\ref{orb_magn_mz}a and b, respectively. 
In the first case, we observe isotropic Friedel oscillations in the induced orbital 
magnetization around the Fe impurity since the spin moment does not break the 
cylindrical symmetry. Similarly to the current density, $m_{l,z}(\vec{r})$ oscillates 
with two characteristic wave lengths $\lambda_\text{F}$ and $\lambda_\text{so}.$
These oscillations decay as $\frac{1}{r}$, which is slower than the induced spin 
magnetization~\cite{Lounis:2012}. Nonetheless, $m_{l,z}(\vec{r})$ is one order of 
magnitude smaller in comparison to the induced spin magnetization density. The net 
orbital and spin magnetizations are $M_{l,z} = -0.58 \,\mu_\text{B}$ and $M_{s,z} 
= 2.11\,\mu_\text{B}$, respectively.
 
For the in-plane orientation depicted in Fig.~\ref{orb_magn_mz}b where the cylindrical 
symmetry is broken, two oscillation wave lengths are also found: $\lambda_\text{F}$ and 
$\lambda_\text{so}$. The orbital magnetization density $m_{l,z}(\vec{r})$, however, 
oscillates around a positive (negative) value for $x < 0$ ($x > 0$). The oscillations 
are less pronounced compared to the case where the impurity has a moment along the 
$z$-axis in Fig.~\ref{orb_magn_mz}a. Nevertheless, the order of magnitude and the 
asymptotic decay of $m_{l,z}(\vec{r})$ at large distances are similar in both 
cases. Furthermore, when the spin moment points in the plane, $m_{l,z}(x,y) = 
-m_{l,z}(-x,y)$ and, therefore, the total induced orbital magnetization sums up 
to zero. The net induced spin magnetization vanishes as well.

\section{Orbital magnetization for different types of impurities}
\label{orb_magn_scatt}

Here, we consider a single impurity with $\vec{S}_{i}\parallel z$-axis and investigate 
the dependence of the induced orbital magnetization on the nature of the magnetic 
impurities. This is achieved by shifting the position of the impurity resonance 
with respect to the Fermi energy of the Rashba electron gas, which represents
different charge and spin states of the impurity. The impurities are 
modeled using a scattering phase shift $\delta_{i}^{\sigma}(\varepsilon)$ that can 
be related to its local density of states $n_{i}(\varepsilon)$ via the Friedel sum 
rule~\cite{Friedel:1958}
\begin{equation}
n_{i}(\varepsilon) = \frac{1}{\pi}\,\frac{\dd\delta_{i}^{\sigma}(\varepsilon)}
{\dd\varepsilon}\quad.
\end{equation}
We focus on $3d$ transition metal impurities for which the local density of states 
has a Lorentzian-like shape~\cite{Lounis:2006,Stepanyuk:1996}. Thus, the scattering 
phase shift can be computed analytically and reads 
\begin{equation}
\delta_{i}^{\sigma}(\varepsilon) = \frac{\pi}{2} + \text{atan}\left(\frac{\varepsilon-
	\varepsilon_{i}^{\sigma}}{\Gamma_{\sigma}}\right)\quad,
	\label{scattering_phase_shift}
\end{equation}
$\varepsilon_{i}^{\sigma}$ being the resonance position for the spin channel $\sigma$ 
and $\Gamma_{\sigma}$ is the resonance width at half maximum. The broadening of the 
impurity states is induced by hybridization with the Rashba electron gas and with other 
substrate electronic states not explicitly being considered. Furthermore, for the $3d$ 
transition metal impurities of interest, the majority spin is fully occupied 
(\textit{i.e.} $\delta_{i}^{\uparrow}(\varepsilon) = \pi$) and does not contribute to the 
bound current density (see Sec.~\ref{Bound_current_tmat}). 

The net orbital magnetization can be obtained by integrating the orbital magnetization 
density computed Eq.~\eqref{poisson_magn2} over the simulation box. Alternatively, we 
can provide an approximate connection between $M_{l,z}$ and the transition matrices (and, 
therefore, with $\delta^{\sigma}_{i}(\varepsilon)$) using the classical formula: 
\begin{equation}
\vec{M}_{l} = \frac{1}{2}\int_{S} \dd\vec{r}\,\vec{r}\times\vec{j}(\vec{r})\quad.
\label{current_orb_magn}
\end{equation}
Starting from Eq.~\eqref{current_mz} and using the asymptotic forms of $G_\text{D}(r,\varepsilon)$  
and $G_\text{ND}(r,\varepsilon)$ for ${r} \rightarrow \infty$ and then performing the spatial 
integral in the previous equation, we find the following approximate expression: 
\begin{equation}
M^\text{app}_{l,z} \propto \text{Re} \int_{0}^{\varepsilon_\text{F}}\dd\varepsilon
\,\frac{\Delta t_{i}(\varepsilon)}{\sqrt{\varepsilon}}\quad.
\end{equation}

\begin{figure}
	\hspace{-5mm}
	\includegraphics[width=0.45\textwidth]{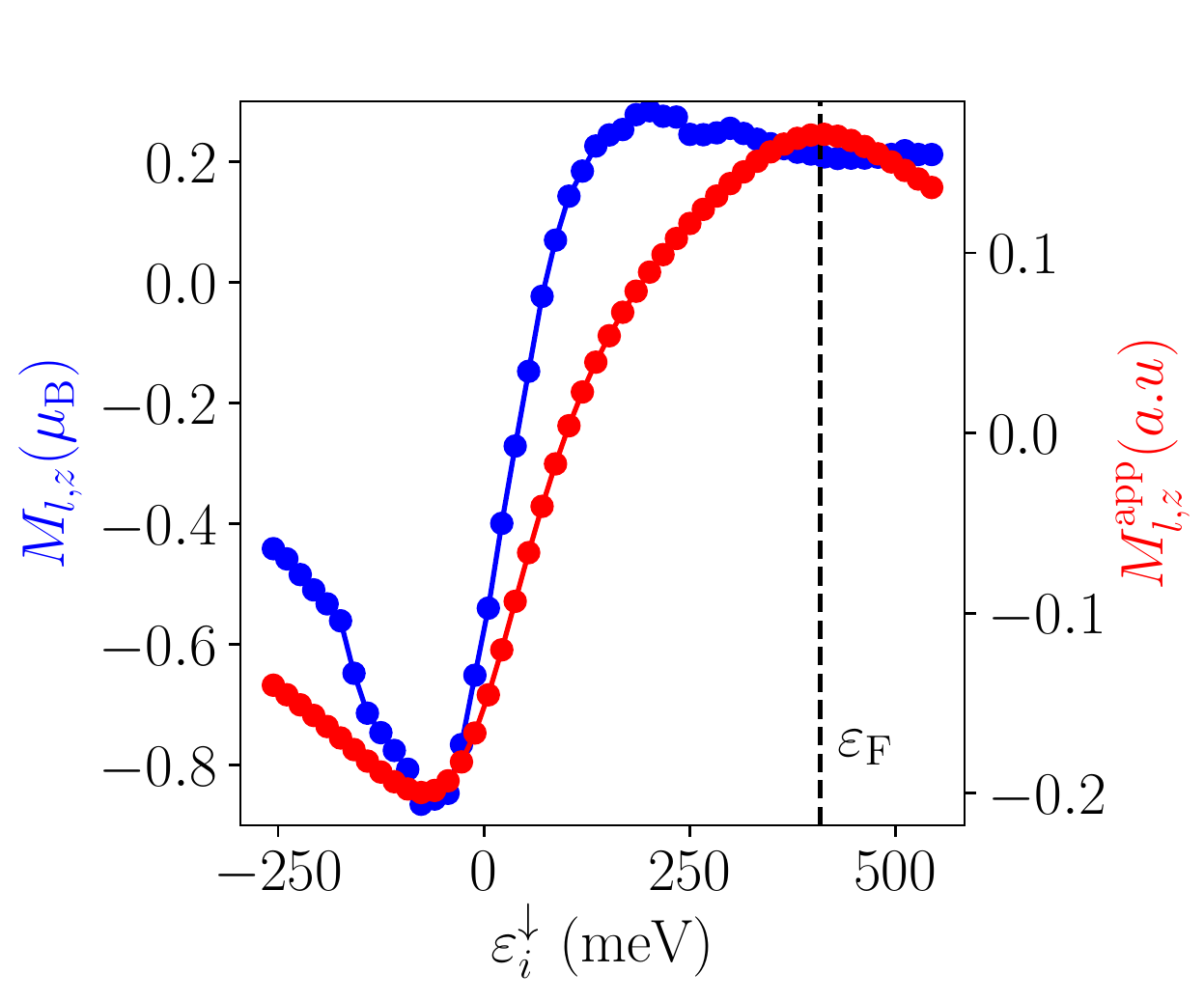}
	\caption{Evolution of $M_{l,z}$ (blue curve) and of $M^\text{app}_{l,z}$ (red curve) 
	as a function of the impurity resonance position $\varepsilon^{i}_{\downarrow}$ 
	(minority spin channel). Both curves have a similar behavior and display a band 
	filling effect. The broadening of the minority spin channel is set to 
	$\Gamma_{\downarrow}= 115$ meV and the Fermi energy $\varepsilon_\text{F} = 410$ 
	meV.} 
	\label{Morb_scatt_matrix}
\end{figure}

In Fig.~\ref{Morb_scatt_matrix}, we show $M_{l,z}$ and $M^\text{app}_{l,z}$ as a function 
of the resonance position of the minority spin channel $\varepsilon_{\downarrow}$.  
$\Delta t_{i}(\varepsilon)$ is computed using the energy dependent scattering phase 
shift given in Eq.~\eqref{scattering_phase_shift}. Both quantities follow the same 
trend and display a step-like feature,
showing a dependence on the valence of the impurity.
Furthermore, since we assumed that the majority impurity resonance is fully occupied, in the limit $\varepsilon_{i}^{\downarrow} 
\rightarrow -\infty$, both resonances become fully occupied and the net orbital magnetization 
vanishes. Moreover, when $\varepsilon_{i}^{\downarrow} \rightarrow +\infty$, $M_{l,z}$ 
vanishes as well since the occupied Rashba states do not hybridize with the impurity states.
This shows that the nature of the impurity has a deep impact on the induced orbital 
magnetization, and the valence of the impurity can be employed to tune its magnitude. 

\section{Orbital magnetization of a trimer on a Rashba electron gas}

\begin{figure*}
	\centering
	\includegraphics[width=1.0\textwidth]{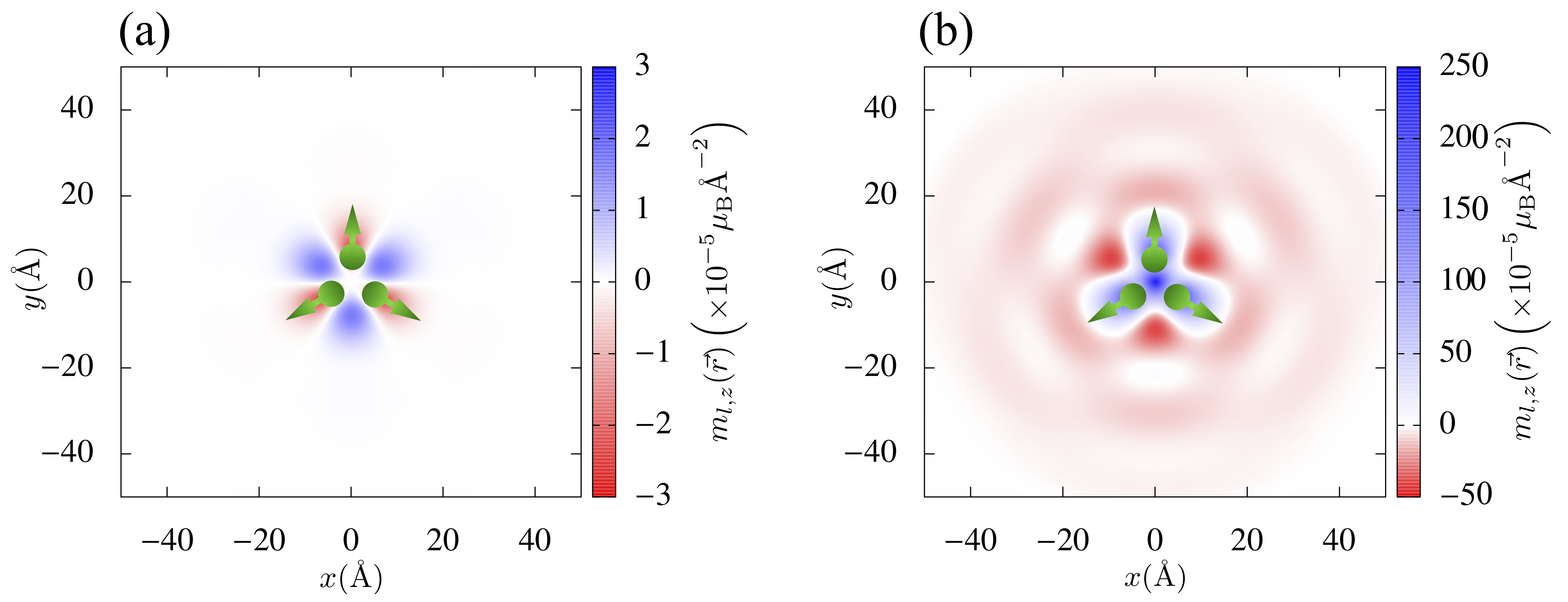}
	\caption{Induced orbital magnetization map for an Fe trimer deposited 
		on a Rashba electron gas with an equilateral triangle geometry. We used 
 	    the same model parameters as in Sec.~\ref{Bound_current_tmat}. The Fe 
 	    impurities are represented by a green sphere located at the origin, while 
 	    their magnetic moment is represented by a green arrow (The Fe impurities 
 	    are considered in the s-wave approximation~\cite{Fiete:2003}).
        The magnetic trimer has an opening angle of $\theta = 60^\circ$ 
        and the azimuthal angles are $\phi_{i} =\{330^{\circ}, 90^{\circ}, 210^{\circ}\}$, 
        respectively.
		(a) In absence of SO interaction, the orbital magnetization remains finite,
	    it follows $C_{3v}$ symmetry and is rather small. (b) In presence of SO interaction, 
	    the orbital magnetization is two orders of magnitude higher in comparison with the
	    previous case and displays a constructive interference at the center 
	    of mass of the equilateral triangle.} 
	\label{orb_magn_trimer}
\end{figure*}

After investigating the emerging orbital magnetization induced by a single magnetic 
impurity, we consider now a more complex nanostructure composed of three Fe atoms 
forming an equilateral triangle centered at the origin. The distance between the Fe 
impurities is $d = \SI{10.42}{\angstrom}$, corresponding to the seventh nearest 
neighbor distance on Au(111) (long distance regime where the s-wave approximation is 
valid).
For this separation the impurity Fe magnetic 
moments are coupled antiferromagnetically~\cite{Juba:2017} leading to a 
non-collinear magnetic state. The ground state without the presence of 
spin-orbit interaction is a N\'eel-state with an angle between the impurity 
spins of $120^{\circ}$. We first begin by omitting the contribution of the SO interaction and assume 
that the moments are non-coplanar with an opening polar angle of $\theta = 60^{\circ}$ 
and an azimuthal angle $\phi_{i} = \{330^{\circ}, 90^{\circ}, 210^{\circ}\}$. The resulting 
orbital magnetization is shown in Fig.~\ref{orb_magn_trimer}a. Even though the SO interaction 
is absent, $m_{l,z}(\vec{r})$ is finite and follows the $C_{3v}$ symmetry of the 
system~\cite{Manuel:2016}. In this case, the current density and, consequently, the 
induced angular momentum have their origin in the non-collinearity of the moments and 
can be traced to the scalar three-spin chirality $\vec{S}_{i}\cdot(\vec{S}_{j}\times\vec{S}_{k})$ 
and its higher-order generalizations (see Appendix~\ref{Derivation_tom}). For that reason, 
we refer to this contribution as \textit{chiral orbital magnetization}. Similarly 
to the single atom case, $m_{l,z}(\vec{r})$ oscillates with two wave lengths 
$(\lambda_\text{F}$ and $\lambda_\text{so})$. The results obtained in 
Fig.~\ref{orb_magn_trimer}a also reveals that the induced net chiral orbital 
magnetization vanishes by symmetry in the simulation box. The net induced spin 
magnetization is however finite $M_{s,z} = -0.16\,\mu_\text{B}$.

\begin{figure}[t]
	\hspace{-10mm}
	\includegraphics[width=0.450\textwidth]{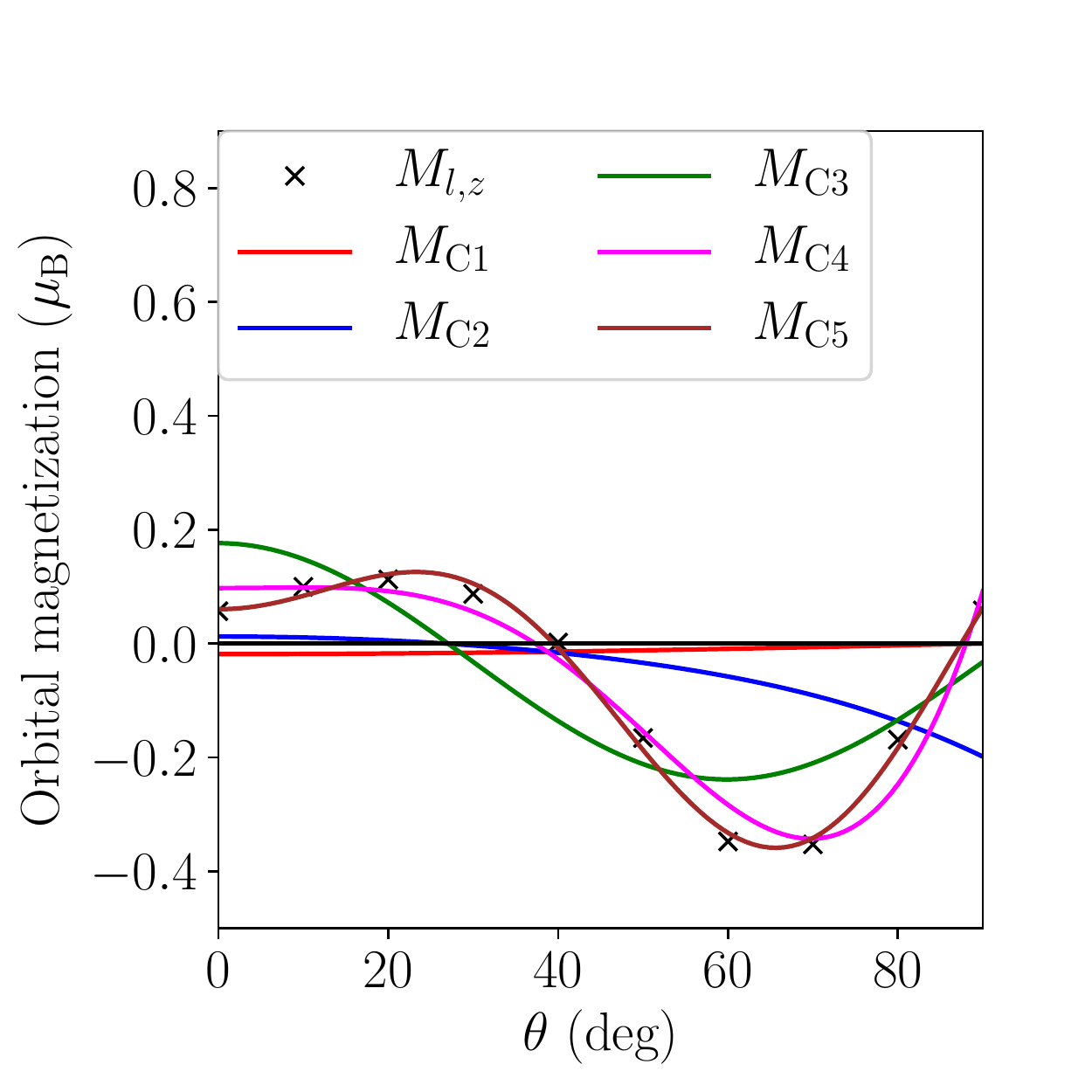}
	\caption{Net orbital magnetization $(M_{l,z})$ as functions of the opening polar 
		angle $\theta$ of an Fe trimer in an equilateral triangle geometry (see 
		Fig.~\ref{orb_magn_trimer}). The black crosses indicate the values of the 
		net orbital magnetization computed using Eq.~\eqref{fourier_transform_morb}. 
		The full curves represents the fits of the net orbital magnetization up to 
		different orders.} 
	\label{orb_magn_trimer_2}
\end{figure}

Including the contribution of the SO interaction, the obtained orbital magnetization
is shown in Fig.~\ref{orb_magn_trimer}b. Similarly to the SO interaction free case, 
$m_{l,z}(\vec{r})$ obeys $C_{3v}$ symmetry and has a ${1}/{r}$ decay, but now with 
values two orders of magnitude larger than the chiral contribution. The oscillation 
in $m_{l,z}(\vec{r})$ are more pronounced, and the constructive interference at the 
center of mass of the triangle gives rise to large values of $m_{l,z}(\vec{r})$. 
The net induced spin and orbital magnetizations are $M_{s,z} = -0.49\,\mu_\text{B}$ and 
$M_{l,z} = -0.35\,\mu_\text{B}$, respectively. In presence of SO interaction, the 
connection between $m_{l,z}(\vec{r})$ and the non-collinear spin texture is more complex. 
It was shown previously in Ref.~\onlinecite{Fabian:2018} that for continuous spin textures, 
the presence of two non-collinear spin moments is enough to influence the orbital 
magnetization. In the limit of fast rotating spin textures, similar terms arise in 
$M_{l,z}$. However and as shown in the following, higher order contributions besides the three-spin chirality
can be of crucial importance and strongly influence its angular dependence.

To study the dependence of the total induced orbital magnetization $M_{l,z}$ (including 
both chiral and SO contributions) on the spin orientation of the impurities, we computed 
it for different opening angles $\theta$. The result is shown in Fig.~\ref{orb_magn_trimer_2}, 
where we notice that $M_{l,z}$ can be rather large, reaching $-0.4\,\mu_\text{B}$ for 
$\theta = 70^\circ$. Then by performing a Born expansion of the Green function and retaining 
only terms up to first order in SO interaction (similarly to Appendix~\ref{Derivation_tom}), 
$M_{l,z}$ can be written as:
\begin{equation}
\begin{split}
M_{l,z}&= \beta_{1}\,\cos\theta + \beta_{2}\, \cos\frac{\gamma}{2} \\ & + 
\beta_{3}\,\sin^{2}\theta\cos\theta + \beta_{4}\, \cos^{2}\frac{\gamma}{2} \\ & 
+ \beta_{5}\, \cos{\gamma}\,\sin^{2}\theta\cos\theta\quad,
\end{split}
\label{f_form_om}
\end{equation}
where $\gamma = \arccos(\vec{S}_{1}\cdot\vec{S}_{2})$ is the angle 
between $\vec{S}_{1}$ and $\vec{S}_{2}$, respectively.
As shown in Appendix~\ref{Derivation_tom}, in absence of SO interaction 
$\beta_{1} = \beta_{2} = \beta_{4} = 0$ and the functional form includes 
only the odd powers of the spin moments. We also show in Fig.~\ref{orb_magn_trimer_2} 
the different fits of the orbital magnetization obtained when truncating 
Eq.~\eqref{f_form_om} at different orders and provide the values of the 
coefficients $\beta_{i}$ in Table~\ref{tab_param_om}. This reveals the 
importance of higher order contributions to capture the right angular 
dependence of $M_{l,z}$ in the entire range of angles taken into account. 
The low order expansions would only be able to reproduce the correct behaviour 
in a small angular window. 

\begin{table}[t]
	\begin{center}
		\begin{tabular}{lllllr}
			\hline
			Parameters       & $\beta_{1}$     & $\beta_{2}$     &  $\beta_{3}$ & $\beta_{4}$ & $\beta_{5}$\\
            \hline
            $M_{\text{C}1}$  & -0.019          & ---             & ---          & ---         & ---    \\
            \hline 
            $M_{\text{C}2}$  &  0.409          & -0.397          & ---          & ---         & ---    \\
            \hline 
            $M_{\text{C}3}$  &  0.242          & -0.065          & -0.843       & ---         & ---    \\
            \hline
            $M_{\text{C}4}$  & -7.653          & -7.385          &  4.817       & 15.136      & ---    \\
            \hline
            $M_{\text{C}5}$  & 1.854           & 2.042           & -2.244       & -3.836      &  1.932 \\
			\hline
		\end{tabular}
		\caption{Parameters used to fit the orbital magnetization up to fifth power of 
			spin impurity moment. $M_{\text{C}i}$ stands for the fit of the orbital magnetization 
			up to the $i^\text{th}$ order. $C$ stands for chirality. The fits are displayed 
			in Fig.~\ref{orb_magn_trimer_2}.}
		\label{tab_param_om}
	\end{center}
\end{table}

\section{Discussions and conclusion}
In this paper, we used a model approach relying on the Rashba Hamiltonian 
to understand the emergence of an induced orbital magnetization when magnetic 
impurities are incorporated into a Rashba electron gas. The magnetic impurities
were described using scattering phase shifts that were either taken to be constant or were obtained with a Lorentzian-like 
shape of the impurity density of states. We computed the dissipationless bound 
currents present in the system, which consist of paramagnetic and diamagnetic-like 
contributions, devising a method applicable to any ensemble of impurities with 
an arbitrary magnetic configuration. Afterwards, we showed analytically that, in presence of a single 
magnetic impurity with its moment parallel to the $z$-axis, a finite orbital 
magnetization arises when time and space inversion symmetries are simultaneously 
broken. The net orbital magnetization was found to be of the order of magnitude 
of its spin counterpart. However, it vanishes by symmetry when the impurity spin 
moment lies in the surface plane.
 
The dependence of the net orbital magnetization on the nature of the impurity
was also addressed. Its magnitude and sign strongly depend on the impurity kind 
(namely, its valence and the location of the impurity resonances with respect to the Fermi energy of the electron gas).
Moreover, we considered a more complex magnetic structure consisting of a magnetic 
trimer in an equilateral geometry. In absence of SO interaction and when the spin 
texture displays a non-vanishing scalar spin chirality, a chiral orbital magnetization 
is observed~\cite{Manuel:2016}. This result was also recovered analytically. When 
the SO interaction is turned on, the dependence of the net orbital magnetization 
on the spin texture is more complex, and higher order powers of the spin chirality 
can be of crucial importance. In this case, the orbital magnetization density was 
two orders of magnitude higher in comparison to the case where the SO interaction is not 
present.

Finally, we foresee the possibility of measuring the total (spin/orbital) induced magnetization at surface utilizing spin-polarized STM~\cite{Meier:2011}. Distinguishing, however, the spin from the orbital magnetization is not trivial. One has to consider the asymptotic behavior and the specific decay of both types of magnetization. Since the magnetization density produces stray fields, measurements exploiting Nitrogen-Vacancy centers might enable their detection~\cite{Toyli:2012}. Moreover, we believe that XMCD is a possible technique to track higher order spin-chiralities, which play a major role for the determination of induced orbital magnetization by altering the topological properties of the orbital magnetization induced by complex magnetic structures such as magnetic skyrmions~\cite{Manuel:2016}.

\textbf{Acknowledgements} 
This work was supported by the European Research Council (ERC) under the European 
Union's Horizon 2020 research and innovation programme (ERC-consolidator grant 681405 
DYNASORE).

\appendix

\section{Single magnetic adatom ground state charge current}
\label{sec:Appendix_A}

In this Appendix, we derive the ground state charge current induced by magnetic 
impurities with a spin moment perpendicular to the plane containing the Rashba 
electron gas (\textit{i.e.} along the $z$-axis). The current operator given in 
Eq.~\eqref{current_op} contains a gradient acting on the Green function. Since 
the cylindrical symmetry of the Rashba electron gas is preserved when the moment 
points out of the plane, we write the gradient in cylindrical coordinates as
\begin{equation}
\vec{\nabla}_{\vec{r}} = {\vec{e}_{r}}\,\frac{\partial}{\partial r} + \frac{1}{r}
\,{\vec{e}_{\theta}}\,\frac{\partial}{\partial \theta}\quad.
\end{equation}
$\vec{e}_{r}$ and $\vec{e}_{\theta}$ are the radial and azimuthal unit vectors, 
respectively. Assuming an impurity located at position $\vec{R}_{i}$, we define the 
gradient accordingly as $\vec{\nabla}_{\vec{r}_{i}}$. Furthermore, the Rashba Green 
function is a matrix in spin space, given by
\begin{equation}
\boldsymbol{G}^\text{R}(\vec{R},\varepsilon+\iu 0^+)=
\left(
\begin{array}{cc}
G_{\text{D}}(R,\varepsilon)             &   -G_{\text{ND}}(R,\varepsilon)\,e^{-\iu\beta}\\
G_\text{ND}(R,\varepsilon)\,e^{\iu\beta}  &    G_\text{D}(R,\varepsilon)\\
\end{array}
\right)\quad,
\end{equation}
where $G_{\text{D}}(R,\varepsilon)$ and $G_{\text{ND}}(R,\varepsilon)$ are given by linear 
combinations of Hankel functions of zero and first order, respectively,
\begin{equation}
\begin{split}
G_\text{D}(R,\varepsilon+\iu 0^+) =& -\frac{\iu m^*}{2\hbar^2(k_++k_-)}\left[\right. k_+\,H_{0}(k_+R+\iu 0^+)\\
&+k_-\,H_0(k_-R +\iu 0^+)\left.\right]\quad ,
\label{g01}
\end{split}
\end{equation}
\begin{equation}
\begin{split}
G_\text{ND}(R,\varepsilon+\iu 0^+) = &-\frac{\iu m^*}{2\hbar^2(k_++k_-)}\left[\right. k_+\,H_1(k_+R +\iu 0^+)\\
&-k_-\,H_1(k_-R +\iu 0^+)\left.\right]\quad .
\label{g02}
\end{split}
\end{equation}
The wave vectors $k_{+}$ and $k_{-}$ are given by $k_+ = k_\text{so} + 
\sqrt{k_\text{so}^2+\frac{2m^*\varepsilon}{\hbar^2}}$ and $k_-=- k_\text{so} 
+ \sqrt{k_\text{so}^2+\frac{2m^*\varepsilon}{\hbar^2}}$ with $k_\text{so} = 
\frac{m^* \alpha_\text{so}}{\hbar^2}$. The gradient of the Rashba Green function 
is given by:
\begin{widetext}
\begin{equation}
\vec{\nabla}_{\vec{r_{i}}}\,\boldsymbol{G}^\text{R}(\vec{r_{i}},\varepsilon+\iu 0^+) =
\begin{pmatrix}
\vec{e}_{r_i}\,\frac{\partial G_\text{D}}{\partial r_{i}} & e^{-\iu\theta_{i}}[-\vec{e}_{r_{i}}\,
\frac{\partial G_\text{ND}}{\partial r_{i}} + \iu\,{\vec{e}_{\theta_{i}}}\,\frac{G_\text{ND}}{r_{i}}]\\
e^{\iu\theta_{i}}[{\vec{e}_{r_{i}}}\,\frac{\partial G_\text{ND}}{\partial r_{i}} + \iu\,{\vec{e}_{\theta_{i}}}\ 
\frac{G_\text{ND}}{r_{i}}] & {\vec{e}_{r_{i}}}\,\frac{\partial G_\text{D}}{\partial r_{i}}
\end{pmatrix},
\label{grad_gf}
\end{equation}
\end{widetext}
From Eq.~\eqref{grad_gf} $\frac{\partial G_\text{D}}{\partial r_{i}}$ and $\frac{\partial 
G_\text{ND}}{\partial r_{i}}$ are needed. This involves first order derivatives of Hankel 
functions which can be computed using recursion:
\begin{equation}
\frac{dH_{n}(x)}{dx} = \left[\frac{n\,H_{n}(x)}{x} - H_{n+1}(x)\right]\quad.
\end{equation}
For the Rashba Green function one needs the derivatives of $H_{0}(x)$ and $H_{1}(x)$:
\begin{eqnarray}
\left\{ \begin{array}{ll}
\frac{dH_{0}(x)}{dx} = -H_{1}(x)\quad,
\\ \nonumber
\frac{dH_{1}(x)}{dx} = \left[\frac{\ H_{1}(x)}{x} - H_{2}(x)\right]\quad.
\end{array}\right.\hspace{5ex}
\end{eqnarray}
After computing $\vec{\nabla}_{\vec{r_{i}}}\,\boldsymbol{G}^\text{R}(\vec{r_{i}},\varepsilon+\iu 0^+)$ 
one can easily access $\vec{\nabla}_{\vec{r_{i}}}\,\boldsymbol{G}(\vec{r_{i}},\varepsilon+\iu 0^+)$ 
via Eq.~\eqref{dyson_eq_gf}, which is employed to compute the expectation value of 
${\vec{\boldsymbol{j}}}$ given in Eq.~\eqref{current_op} via: 
\begin{equation}
\vec{j}(\vec{r}) = \int_{-\infty}^{\varepsilon_\text{F}}\dd\varepsilon\,\text{Tr}
\,{\vec{\boldsymbol{j}}}\,\boldsymbol{G}(\vec{r},\varepsilon)\quad,
\label{current_ex}	
\end{equation}
where the trace is taken over the spin degree of freedom.

\section{Paramagnetic charge current without SO interaction}
\label{Derivation_tom}

Here, we derive the connection between the chiral orbital magnetization and the scalar 
chirality (and the spin texture in general) up to the fifth order. We consider that the 
spin-orbit interaction is zero (\textit{i.e.} $\alpha_\text{so} = 0$), thus the Rashba 
Green function becomes spin diagonal. Then we perform a Born expansion of  Eq.~\eqref{scatter_path} 
as
\begin{equation}
\begin{split}
\boldsymbol{G}(\vec{r},\vec{r}^{\,\prime},\varepsilon) =\, 
&\boldsymbol{G}^\text{R}(\vec{r},\vec{r}^{\,\prime},\varepsilon)
\,+ \boldsymbol{G}^{(1)}(\vec{r},\vec{r}^{\,\prime},\varepsilon)\,+\\
&\boldsymbol{G}^{(2)}(\vec{r},\vec{r}^{\,\prime},\varepsilon)\,+
\boldsymbol{G}^{(3)}(\vec{r},\vec{r}^{\,\prime},\varepsilon)\,+\\
&\boldsymbol{G}^{(4)}(\vec{r},\vec{r}^{\,\prime},\varepsilon)\,+
\boldsymbol{G}^{(5)}(\vec{r},\vec{r}^{\,\prime},\varepsilon)\,+ ....
\end{split}
\label{dyson_expansion}
\end{equation}
The different elements of the expansion $G^{(i)}(\vec{r},\vec{r}^{\,\prime},\varepsilon)$ are
\begin{equation}
\boldsymbol{G}^{(1)}(\vec{r},\vec{r}^{\,\prime},\varepsilon) =  
\sum_{i} \boldsymbol{G}^\text{R}(\vec{r},\vec{r_{i}},\varepsilon)\,  
\boldsymbol{t}_{i}(\varepsilon)\, 
\boldsymbol{G}^\text{R}(\vec{r_{i}},\vec{r}^{\,\prime},\varepsilon)\quad ,
\end{equation}
\begin{equation}
\begin{split}
\boldsymbol{G}^{(2)}(\vec{r},\vec{r}^{\,\prime},\varepsilon) = 
\sum_{ij}& \boldsymbol{G}^\text{R}(\vec{r},\vec{r_{i}},\varepsilon)\,  
\boldsymbol{t}_{i}(\varepsilon)\, \boldsymbol{G}^\text{R}(\vec{r_{i}},\vec{r_{j}},\varepsilon)\,\\ &\boldsymbol{t}_{j}(\varepsilon)\,\boldsymbol{G}^\text{R}(\vec{r_{j}},\vec{r}^{\,\prime},\varepsilon)\quad ,
\end{split}
\end{equation}
\begin{equation}
\begin{split}
\boldsymbol{G}^{(3)}(\vec{r},\vec{r}^{\,\prime},\varepsilon) = 
\sum_{ijk}& \boldsymbol{G}^\text{R}(\vec{r},\vec{r_{i}},\varepsilon)\,  
\boldsymbol{t}_{i}(\varepsilon)\, \boldsymbol{G}^\text{R}(\vec{r_{i}},\vec{r_{j}},\varepsilon)
\, \boldsymbol{t}_{j}(\varepsilon)\\&\boldsymbol{G}^\text{R}(\vec{r_{j}},\vec{r}_{k},\varepsilon) \boldsymbol{t}_{k}(\varepsilon)\,\boldsymbol{G}^\text{R}(\vec{r_{k}},\vec{r}^{\,\prime},\varepsilon)\quad ,
\end{split}
\end{equation}
\begin{equation}
\begin{split}
\boldsymbol{G}^{(4)}(\vec{r},\vec{r}^{\,\prime},\varepsilon)=  
\sum_{ijkm}& \boldsymbol{G}^\text{R}(\vec{r},\vec{r_{i}},\varepsilon)\,  
\boldsymbol{t}_{i}(\varepsilon)\, \boldsymbol{G}^\text{R}(\vec{r_{i}},\vec{r_{j}},\varepsilon)
\, \boldsymbol{t}_{j}(\varepsilon)\\&\boldsymbol{G}^\text{R}(\vec{r_{j}},\vec{r}_{k},\varepsilon) \boldsymbol{t}_{k}(\varepsilon)\,\boldsymbol{G}^\text{R}(\vec{r_{k}},\vec{r}_{m},\varepsilon)
\\&\boldsymbol{t}_{m}(\varepsilon)\,\boldsymbol{G}^\text{R}(\vec{r}_{m},\vec{r}^{\,\prime},\varepsilon)\quad,
\end{split}
\end{equation}
\begin{equation}
\begin{split}
\boldsymbol{G}^{(5)}(\vec{r},\vec{r}^{\,\prime},\varepsilon)=  
\sum_{ijk}& \boldsymbol{G}^\text{R}(\vec{r},\vec{r_{i}},\varepsilon)\,  
\boldsymbol{t}_{i}(\varepsilon)\,\boldsymbol{G}^\text{R}(\vec{r_{i}},\vec{r_{j}},\varepsilon)
\, \boldsymbol{t}_{j}(\varepsilon)\\&\boldsymbol{G}^\text{R}(\vec{r}_{j},\vec{r}_{k},\varepsilon)
\,\boldsymbol{t}_{k}(\varepsilon)\,\boldsymbol{G}^\text{R}(\vec{r_{k}},\vec{r}^{\,\prime},\varepsilon)
\,\boldsymbol{t}_{m}(\varepsilon)\\&\boldsymbol{G}^\text{R}(\vec{r}_{m},\vec{r}^{\,\prime},\varepsilon)
\,\boldsymbol{t}_{l}(\varepsilon)\,\boldsymbol{G}^\text{R}(\vec{r}_{l},\vec{r}^{\,\prime},\varepsilon)\quad.
\end{split}
\end{equation}

In absence of spin-orbit the interaction, the current operator given in Eq.~\eqref{current_op} 
contains only the paramagnetic part and reduces to $\frac{e\hbar}{2m^{*}\iu}\, \lim_{\vec{r}^{\,\prime} 
	\rightarrow \vec{r}}(\vec{\nabla}_{\vec{r}} 
- \vec{\nabla}_{{\vec{r}}^{\,\prime}})$. 
Therefore, the current density is given by
\begin{equation}
\vec{j}(\vec{r}) = \vec{j}^{\,(3)}(\vec{r})  + \vec{j}^{\,(5)}(\vec{r})\quad,
\end{equation}
where due to the cyclic properties of the trace, only the odd powers of the expansion 
contribute to the current (the first order vanishes by symmetry). Furthermore, since 
$\boldsymbol{G}^\text{R}(\vec{r},\vec{r}^{\,\prime},\varepsilon)$ is spin diagonal it 
can be taken out of the trace and $\vec{j}^{\,(3)}(\vec{r})$ is simply given by: 
\begin{equation}
\begin{split}
\vec{j}^{\,(3)}(\vec{r})  \propto  \sum_{ijk}&\vec{\nabla}_{\vec{r}}
\,\boldsymbol{G}^\text{R}(\vec{r},\vec{r_{i}},\varepsilon)
\,\boldsymbol{G}^\text{R}(\vec{r_{i}},\vec{r_{j}},\varepsilon)
\,\boldsymbol{G}^\text{R}(\vec{r_{j}},\vec{r}_{k},\varepsilon)\\ 
&\boldsymbol{G}^\text{R}(\vec{r_{k}},\vec{r},\varepsilon)
\,\text{Tr}\,\big{[}\boldsymbol{t}_{i}(\varepsilon)
\,\boldsymbol{t}_{j}(\varepsilon)\,\boldsymbol{t}_{k}(\varepsilon)\\ 
&- \boldsymbol{t}_{k}(\varepsilon)\,\boldsymbol{t}_{j}(\varepsilon)
\,\boldsymbol{t}_{i}(\varepsilon) \big{]}\quad,
\end{split}
\label{current_3rd}
\end{equation}
while the fifth order contribution reads, 
\begin{equation}
\begin{split}
\vec{j}^{\,(5)}(\vec{r}) \propto \sum_{ijkml}&\vec{\nabla}_{\vec{r}}
\,\boldsymbol{G}^\text{R}(\vec{r},\vec{r_{i}},\varepsilon)
\,\boldsymbol{G}^\text{R}(\vec{r_{i}},\vec{r_{j}},\varepsilon)
\,\boldsymbol{G}^\text{R}(\vec{r_{j}},\vec{r}_{k},\varepsilon)\\ 
&\boldsymbol{G}^\text{R}(\vec{r_{k}},\vec{r}_{m},\varepsilon)\,
\boldsymbol{G}^\text{R}(\vec{r}_{m},\vec{r}_{l},\varepsilon)\,
\boldsymbol{G}^\text{R}(\vec{r_{l}},\vec{r},\varepsilon)\,\\
&\text{Tr}\,\big{[}\boldsymbol{t}_{i}(\varepsilon)
\,\boldsymbol{t}_{j}(\varepsilon)\,\boldsymbol{t}_{k}(\varepsilon)
\,\boldsymbol{t}_{m}(\varepsilon)\,\boldsymbol{t}_{l}(\varepsilon)-\\ 
&\boldsymbol{t}_{l}(\varepsilon)\,\boldsymbol{t}_{m}(\varepsilon)
\,\boldsymbol{t}_{k}(\varepsilon)\,\boldsymbol{t}_{j}(\varepsilon)
\,\boldsymbol{t}_{i}(\varepsilon)\big{]}\quad.
\end{split}
\label{current_5th}
\end{equation}
Further simplifications can be made to the expressions given in Eqs.~\eqref{current_3rd} 
and \eqref{current_5th}, considering that 
\begin{equation}
\boldsymbol{t}_{i} = \frac{t^{\uparrow}_{i}+t^{\downarrow}_{i}}{2} + \frac{t^{\uparrow}_{i}-t^{\downarrow}_{i}}{2}\,\vec{\boldsymbol{\sigma}}\cdot\vec{S}_{i}\quad, 
\end{equation}
and using the properties of the Pauli matrices, $\vec{j}^{\,(3)}(\vec{r})$ simplifies to 
\begin{equation}
\vec{j}^{\,(3)}(\vec{r}) = \sum_{ijk} \vec{f}_{3}(\vec{r})
\,\vec{S}_{i}\cdot(\vec{S}_{j}\times\vec{S}_{k})\quad,
\label{current_3rd_final}
\end{equation}
where $\vec{f}_{3}(\vec{r})$ is given by
\begin{equation}
\begin{split}
\vec{f}_{3}(\vec{r}) =-\frac{2e\hbar}{\pi m^{*}} \text{Im}&\,\vec{\nabla}_{\vec{r}}\,\boldsymbol{G}^\text{R}(\vec{r},\vec{r_{i}},\varepsilon)
\,\boldsymbol{G}^\text{R}(\vec{r_{i}},\vec{r_{j}},\varepsilon)\\
&  \boldsymbol{G}^\text{R}(\vec{r_{j}},\vec{r}_{k},\varepsilon)
\,\boldsymbol{G}^\text{R}(\vec{r_{k}},\vec{r},\varepsilon)\quad.
\end{split}
\end{equation}
The fifth order contribution can be also simplified to: 
\begin{equation}
\begin{split}
\vec{j}^{\,(5)}(\vec{r}) = \sum_{ijkml} \vec{f}_{5}(\vec{r})\,
&(\vec{S}_{i}\cdot \vec{S}_{j})\,\left[\vec{S}_{k}\cdot(\vec{S}_{m}\times\vec{S}_{l})\right]\quad,
\end{split}
\label{current_5th_final}
\end{equation}
where $\vec{f}_{5}(\vec{r})$ reads
\begin{equation}
\begin{split}
\vec{f}_{5}(\vec{r}) = & -\frac{4e\hbar}{\pi m^{*}}\text{Im}\,
\vec{\nabla}_{\vec{r}}\,\boldsymbol{G}^\text{R}(\vec{r},\vec{r_{i}},\varepsilon)
\,\boldsymbol{G}^\text{R}(\vec{r_{i}},\vec{r_{j}},\varepsilon)
\,\boldsymbol{G}^\text{R}(\vec{r_{j}},\vec{r}_{k},\varepsilon)\\ 
&\boldsymbol{G}^\text{R}(\vec{r_{k}},\vec{r}_{m},\varepsilon)\,
\boldsymbol{G}^\text{R}(\vec{r}_{m},\vec{r}_{l},\varepsilon)\,
\boldsymbol{G}^\text{R}(\vec{r_{l}},\vec{r},\varepsilon)\quad.
\end{split}
\end{equation}
The previous equations show that in absence of the spin-orbit interaction the 
induced bound currents (\textit{i.e.} orbital magnetization) can be expanded as 
a function of the odd powers of the spin chirality.

\bibliography{mylib_paper.bib}

\end{document}